\documentclass[prd,preprintnumbers,nofootinbib,floatfix]{revtex4} 
\usepackage{graphicx} 
\usepackage{amsmath}
\usepackage{amsfonts,amsbsy}
\usepackage{amssymb}
\usepackage{appendix}

\def\be{\begin{equation}}
\def\ee{\end{equation}}
\def\bea{\begin{eqnarray}}
\def\eea{\end{eqnarray}}

\def\eq#1{{Eq.~(\ref{#1})}}
\def\fig#1{{Fig.~\ref{#1}}}

\def\beq{\begin{equation}}
\def\eeq{\end{equation}}
\def\bea{\begin{eqnarray}}
\def\eea{\end{eqnarray}}
\def\eq#1{{Eq.~(\ref{#1})}}
\def\fig#1{{Fig.~\ref{#1}}}

\def\cpc#1#2#3  {{Computer\ Phys.\ Comm.\ }  {\bf#1}, #2 (#3)}
\def\err#1#2#3  {{\it Erratum }              {\bf#1}, #2 (#3)}
\def\epjc#1#2#3 {{Eur. Phys. J. C }          {\bf#1}, #2 (#3)}
\def\dum#1#2#3  {{~}                         {\bf#1}, #2 (#3)}
\def\ib#1#2#3   {{\it ibid. }                {\bf#1}, #2 (#3)}
\def\jcp#1#2#3  {{J.\ Comput.\ Phys.\ }      {\bf#1}, #2 (#3)}
\def\jetpl#1#2#3 {{\rm JETP Lett.}           {\bf#1}, #2 (#3)}
\def\jhep#1#2#3 {{JHEP }                     {\bf#1}, #2 (#3)}
\def\ijmp#1#2#3 {{Int.\ J.\ Mod.\ Phys.\ }   {\bf#1}, #2 (#3)}
\def\jpg#1#2#3  {{J.\ Phys.\ G }             {\bf#1}, #2 (#3)}
\def\mpl#1#2#3  {{Mod.\ Phys.\ Lett.\ }      {\bf#1}, #2 (#3)}
\def\mpla#1#2#3 {{Mod.\ Phys.\ Lett.\ A }    {\bf#1}, #2 (#3)}
\def\ncim#1#2#3 {{Nuovo Cimento }            {\bf#1}, #2 (#3)}
\def\np#1#2#3   {{Nucl.\ Phys.\ }            {\bf#1}, #2 (#3)}
\def\npb#1#2#3  {{Nucl.\ Phys.\ B}           {\bf#1}, #2 (#3)}
\def\pan#1#2#3  {{Phys.\ At.\ Nuclei }       {\bf#1}, #2 (#3)}
\def\plb#1#2#3  {{Phys.\ Lett.\ B }          {\bf#1}, #2 (#3)}
\def\prep#1#2#3 {{Phys.\ Rep.\ }             {\bf#1}, #2 (#3)}
\def\prd#1#2#3  {{Phys.\ Rev.\ D }           {\bf#1}, #2 (#3)}
\def\prl#1#2#3  {{Phys.\ Rev.\ Lett.\ }      {\bf#1}, #2 (#3)}
\def\ptp#1#2#3  {{Prog.\ Theor.\ Phys.\ }    {\bf#1}, #2 (#3)}
\def\ps#1#2#3   {{Physica Scripta }          {\bf#1}, #2 (#3)}
\def\rmp#1#2#3  {{Rev.\ Mod.\ Phys.\ }       {\bf#1}, #2 (#3)}
\def\rpp#1#2#3  {{Rep.\ Prog.\ Phys.\ }      {\bf#1}, #2 (#3)}
\def\sa#1#2#3   {{Sci. Acta}                 {\bf#1}, #2 (#3)}
\def\sjnp#1#2#3 {{Sov.\ J.\ Nucl.\ Phys.\ }  {\bf#1}, #2 (#3)}
\def\spj#1#2#3  {{Sov.\ Phys.\ JETP }        {\bf#1}, #2 (#3)}
\def\spjl#1#2#3 {{Sov.\ JETP Lett.\ }        {\bf#1}, #2 (#3)}
\def\spu#1#2#3  {{Sov.\ Phys.-Usp.\ }        {\bf#1}, #2 (#3)}
\def\yaf#1#2#3  {{Yad.\ Fiz.\ }              {\bf#1}, #2 (#3)}
\def\zp#1#2#3   {{Zeit.\ Phys.\ }            {\bf#1}, #2 (#3)}
\def\zpc#1#2#3  {{Z.\ Phys.\ C }             {\bf#1}, #2 (#3)}

\begin{document}

\title{\bf Prompt double $J/\psi$ production in proton-proton collisions at the LHC}


\author{S. P. Baranov$^{1}$ and Amir H. Rezaeian$^{2,3}$}
\affiliation{ 
$^1$P. N. Lebedev Institute of Physics, Leninsky prosp. 53, Moscow 117924, Russia\\
$^2$Departamento de F\'\i sica, Universidad T\'ecnica
Federico Santa Mar\'\i a, Avda. Espa\~na 1680,
Casilla 110-V, Valparaiso, Chile\\
$^3$Centro Cient\'\i fico Tecnol\'ogico de Valpara\'\i so (CCTVal), Universidad T\'ecnica
Federico Santa Mar\'\i a, Casilla 110-V, Valpara\'\i so, Chile
}

\begin{abstract}
We provide a detailed study of prompt double $J/\psi$ production within the non-relativistic QCD (NRQCD)  framework in  proton-proton collisions at the LHC. We confront the recent LHC data with the results obtained at leading-order (LO) in the NRQCD framework within two approaches of the collinear factorization and the $k_T$-factorization. We show that the LHCb data are consistent with the $k_T$-factorized LO NRQCD results. We show that the full LO NRQCD formalism cannot describe the recent CMS data, with about one order of magnitude discrepancy. If the CMS data are confirmed, this indicates rather large higher-order corrections for prompt double $J/\psi$ production. We provide various predictions which can further test the NRQCD-based approach at the LHC in a kinematic region that LO contributions dominate. We also investigate long-range in rapidity double $J/\psi$ correlations. We found no evidence of a ridge-like structure for double $J/\psi$ production in proton-proton collisions at the LHC up to subleading $\alpha_s^6$  accuracy. 
\end{abstract}

\maketitle

\section{Introduction}

The nonrelativistic QCD (NRQCD) factorization \cite{nrqcd} is an effective theory for description of the heavy quarkonium production \cite{heavy-c}. In the NRQCD, the production and decay of heavy quarkonium factorizes into two stages, a heavy quark-antiquark pair is first created perturbatively at short distances, calculated by expansion in the strong-coupling constant $\alpha_s$, and then nonperturbatively evolves into quarkonium at long distance. The last stage is described via a universal long-distance matrix elements which are strongly ordered in size by relative velocity $v$ (of the heavy quark) scaling rules \cite{sum}. In the NRQCD framework, the heavy quark-antiquark pairs may appear both as color singlet (CS) and color octet (CO) states. The latter gives rise to the CO mechanism. It is still an open question to understand to what degree the CO mechanism plays a role in quarkonium production. In this paper, we investigate if the current LHC data for prompt double $J/\Psi$ can provide any extra information to address this question.

During last decade there has been tremendous progress in systematic studies of heavy quarkonium production in the NRQCD framework, for a recent review see Ref.\,\cite{heavy-c}.  In particular, it was recently shown that the next-to-leading-order (NLO) calculations for single heavy quarkonium production provide a good description of the experimental data at the LHC and Tevatron \cite{nlo-s}. In contrast, the prompt double heavy quarkonium production is a more complicated process, and at the moment full NLO calculation for this process in the NRQCD framework is not yet available.
Recently He and Kniehl in a nice paper \cite{prl} showed that the full LO up to $\mathcal{O}(\alpha_s^4)$ accuracy  for the prompt double $J/\psi$ production, calculated in the collinearly factorized NRQCD formalism has sizable discrepancies with the recent LHC data. In this paper, we analyse the recent LHC data measured by the LHCb \cite{lhcb} and the CMS \cite{cms} collaborations for the prompt double $J/\psi$ within the LO NRQCD framework but in the $k_T$-factorization approach. The main advantage of the $k_T$-factorization \cite{sg,sg2,sg3} over the collinear factorization \cite{colin} approch is that one can effectively incorporate some of small-x dynamics via initial-state gluon radiations. This is important given the fact that at the LHC, Bjorken $x$ can be small (approximately $x\leq 10^{-2}$ in all experimental data considered here). Here we also extend the previous studies by quantifying various theoretical uncertainties including uncertainties associated to the unintegrated gluon density and our freedom to choose different values for the factorization/renormalization scale. Having quantified all theoretical uncertainties, we examine the importance of CS and CO prescription of the LHC data for double $J/\psi$  production. We show that within the theoretical errors, the  LHCb data are consistent with the LO NRQCD results obtained in the $k_T$-factorization approach by including only CS contribution, while there is more than one order of magnitude discrepancy between the full LO NRQCD results (including both the CS and the CO contributions) and the CMS data both in the collinear factorization and the $k_T$-factorization approaches.  Note that the LHCb and the CMS kinematic coverage for the measurments of prompt double $J/\psi$ production are complementary to each other. Therefore, if the CMS data are confirmed, this indicates rather large higher-order corrections for prompt double $J/\psi$ production, see also Refs.\,\cite{cms0,cms1,cms2,cms3}. We provide various predictions which can further test the NRQCD-based approach at the LHC in a kinematic region that LO contributions dominate.

The discovery of the so-called ridge phenomenon, namely the near-side  di-hadron
correlations which extend to a large pseudorapidity separation, in high-multiplicity events selection in both  proton-proton
and  proton(duetron)-nucleus collisions at the LHC and RHIC  \cite{exp-pp,exp-pa1,exp-pa2,exp-pa3,exp-pa4,exp-pa5,exp-pa6} triggered an on-going debate about the underlying dynamics of such correlations, see for example Refs.\,\cite{ridge0,ridge1,ridge2,ridge3,ridge33,ridge4}. 
A similar ridge-type structure has also been observed in heavy-ion collisions at RHIC and
the LHC and is understood as a phenomenon related to hydrodynamical behavior of the quark-gluon-plasma \cite{exp-geo1,exp-geo2}. 
Nevertheless, it is unknown how/why a small system
like the one produced in proton-proton collisions  can/should exhibit a
hydro type behavior. It is therefore still an open question to understand the true nature of the observed ridge phenomenon, whether it is mainly due to initial-state or final-state physics. Another interesting question is whether the observed ridge phenomenon is universal for all two-particle correlations \cite{2gamma,ridge2-amir} like for example $J/\psi$-pair correlations. In this paper, we also study long-range in rapidity near-side and away-side double $J/\psi$ correlations. We investiage if within the NRQCD framework upto LO, there is any ridge-like structure for $J/\psi$ pair production in proton-proton collisions at the LHC.

This paper is organized as follows: In Sec. II we introduce the main formalism and setup for calculating the cross-section
of inclusive prompt double $J/\Psi$ production in proton-proton collisions within the NRQCD framework. We then present our main results and compare with the recent LHC data in Sec. III. We summarize our main results in Sec. IV.

\section{Theoretical formalism}
In the NRQCD approach, the cross-section of prompt double $J/\psi$ at high-energy can be written in the following general factorization form, 
\begin{equation} \label{e1}
d\sigma\left(p+p\to J/\psi J/\psi+ X\right) = 
 \sum_{m,n}\,\int 
\mathcal{F}_g(x_1, k_{1T})\otimes \mathcal{F}_g(x_2, k_{2T})\otimes
d\sigma\left(g+g\to c\bar{c}\left(n\right)+ c\bar{c}\left(m\right)\right) \langle O^{J/\Psi}_n\rangle \langle O^{J/\Psi}_m\rangle,   
\end{equation}
where $\mathcal{F}_g$ is unintegrated gluon density of the projectile (or target) proton with the longitudinal momentum fraction
$x_{1,2}$ (carried by a parton with respect to the beam proton) and transverse momentum $k_{1T}$ (or $k_{2T}$). The convolution symbol stands for integral phase factors. $d\sigma\left(g+g\to c\bar{c}\left(n\right)+ c\bar{c}\left(m\right)\right)$ is the short-distance coefficients of double charm-pair production in the gluon-gluon (gg) fusion, calculable in powers of $\alpha_s$.  The quark-pair may appear in any Fock state $n,m=^{2S+1}L_j$, both as color singlet (CS) and color octet (CO) states. The heavy-quark pair then evolve nonperturbatively into $J/\Psi$.   
In the case of CO states, extra soft gluon emissions are needed in order to bring the $c\bar{c}$ quantum numbers to the $J/\psi$ state.
The long-distance nature of the heavy quarkonium is factored out into the NRQCD universal long-distance matrix elements (LDMEs) denoted by $ \langle O^{J/\Psi}_n\rangle$,  and are
assumed to obey certain hierarchy in powers of the relative quark velocity $v$ \cite{sum} in the bound state
(this is in a full analog to the classical multipole radiation theory). 
The leading contribution is still due to states with exact $J/\psi$ quantum
numbers (color-singlet) which require no further soft emission. In order to match with the quantum numbers of the final state mesons, one has to project the production amplitude to a proper spin and color state.  Note that the CS contribution dominates for small values of transverse momentum of $J/\Psi$ while the CO contribution are expected to take over at large transverse momentum of  $J/\Psi$ \cite{cs-pt1,cs-pt2,cs-pt3,prl}. This general behaviour is also seen in our results for the prompt double $J/\Psi$ production, see Sec. III. 

\begin{figure}[t]                                       
\includegraphics[width=5 cm] {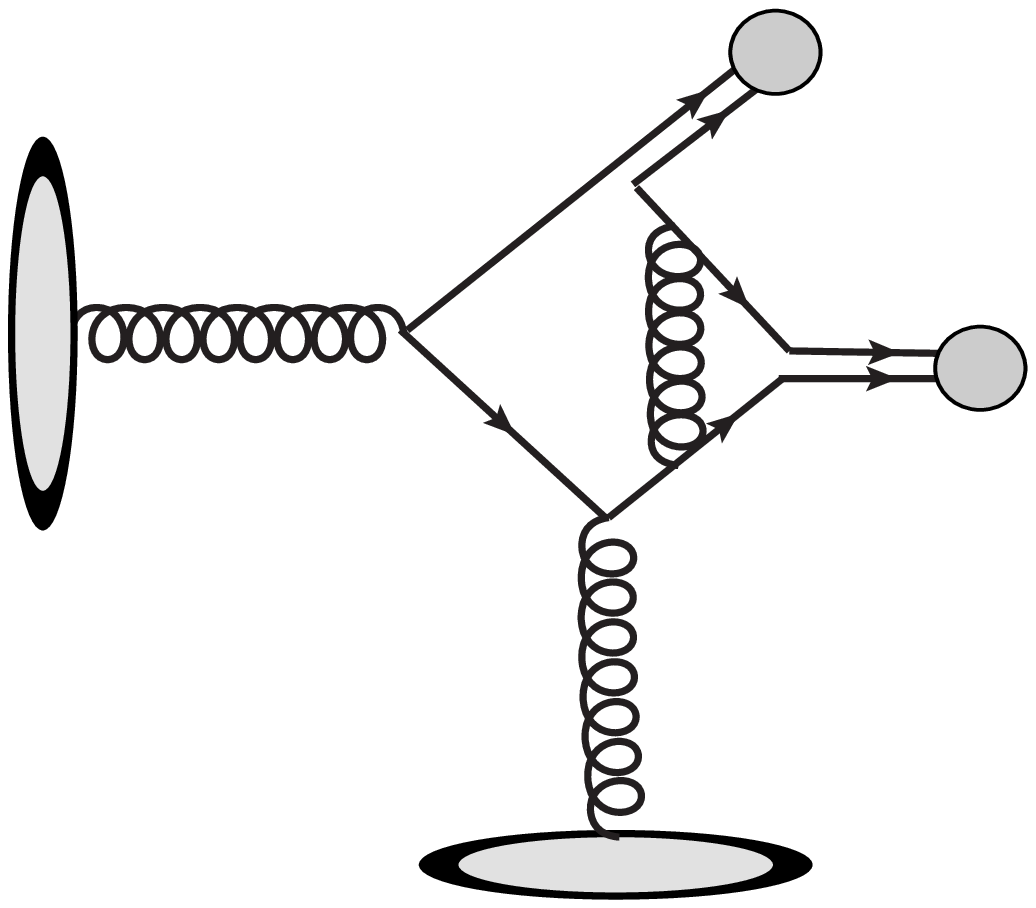}              
\includegraphics[width=10 cm] {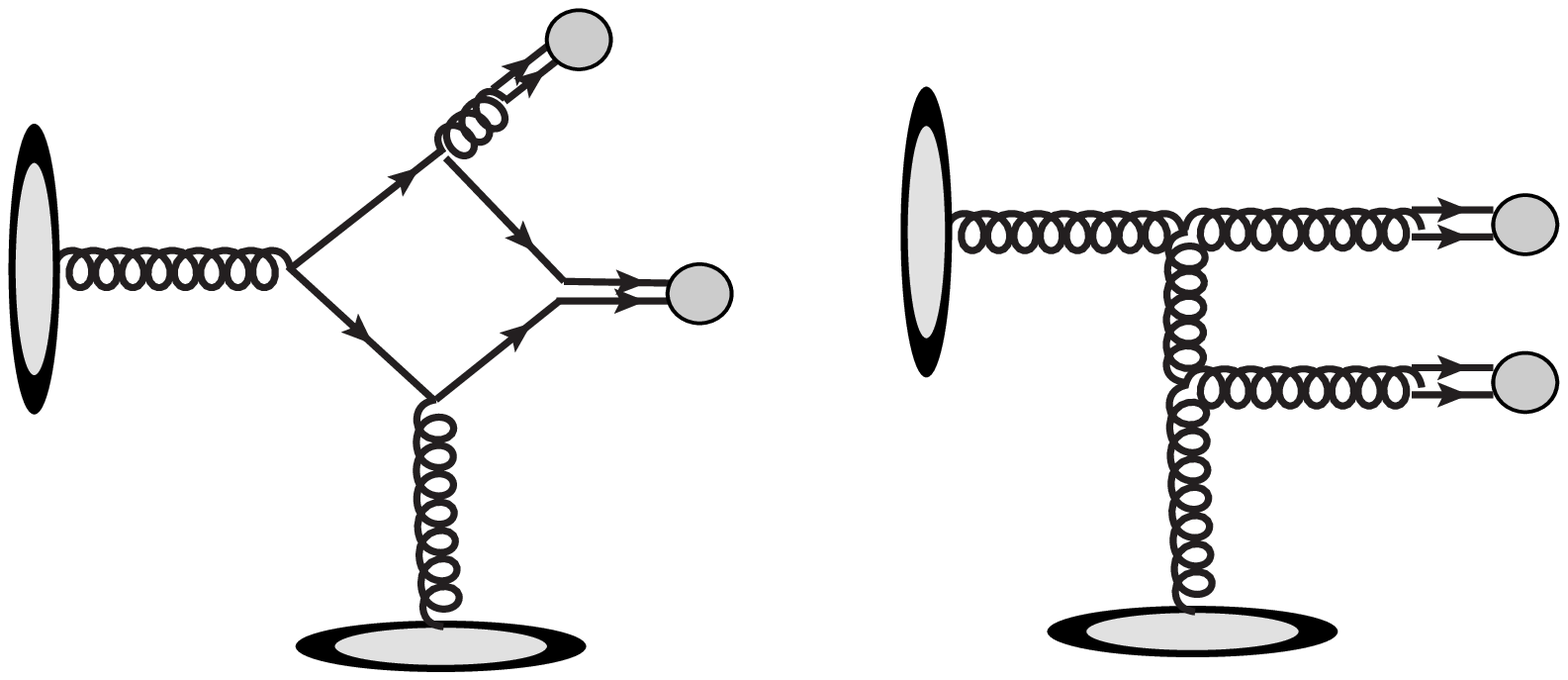}                     
\caption{Typical diagrams contributing to the prompt double $J/\Psi$ production at leading-order $\mathcal{O}(\alpha_s^4)$ in proton-proton collisions. The left panel show the typical color-singlet while the other two diagrams (right) show the typical color-octet diagrams. }
\label{f-di}
\end{figure}

For heavy quarkonium production, the light $q\bar{q}$  contributions are negligible. Moreover, at the LHC energies, one probes the small-x region of parton distribution where gluons dominate.  At leading-order $\mathcal{O}(\alpha_s^4)$, the gluon-gluon fusion subprocess leads to the so-called box diagrams \cite{old2,old0,Scott,Hmery,old1}. Typical diagrams of such processes in the CS and the CO channels are shown in \fig{f-di}.

\begin{figure}[t]                                       
\includegraphics[width=8 cm] {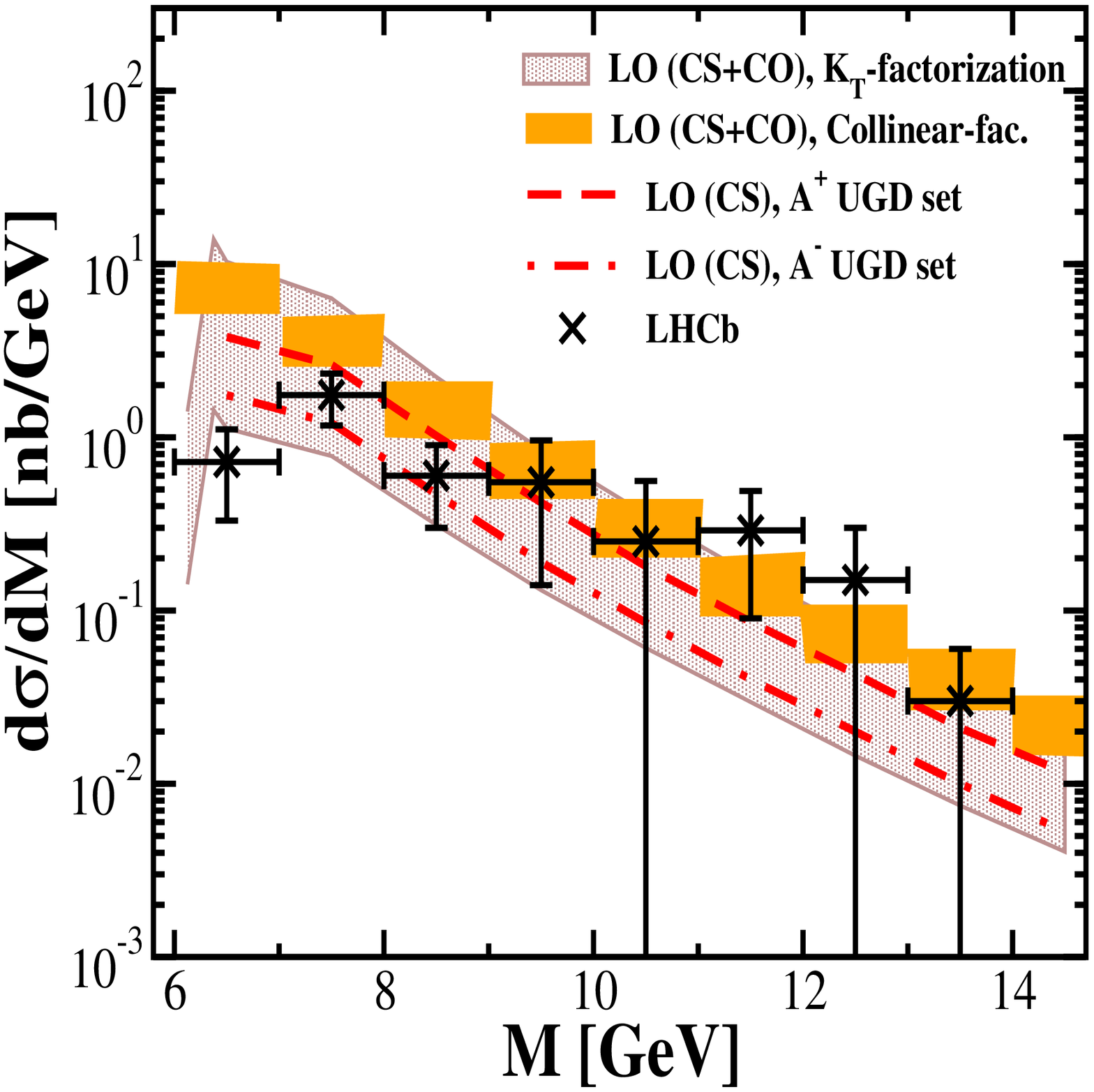}  
\includegraphics[width=8 cm] {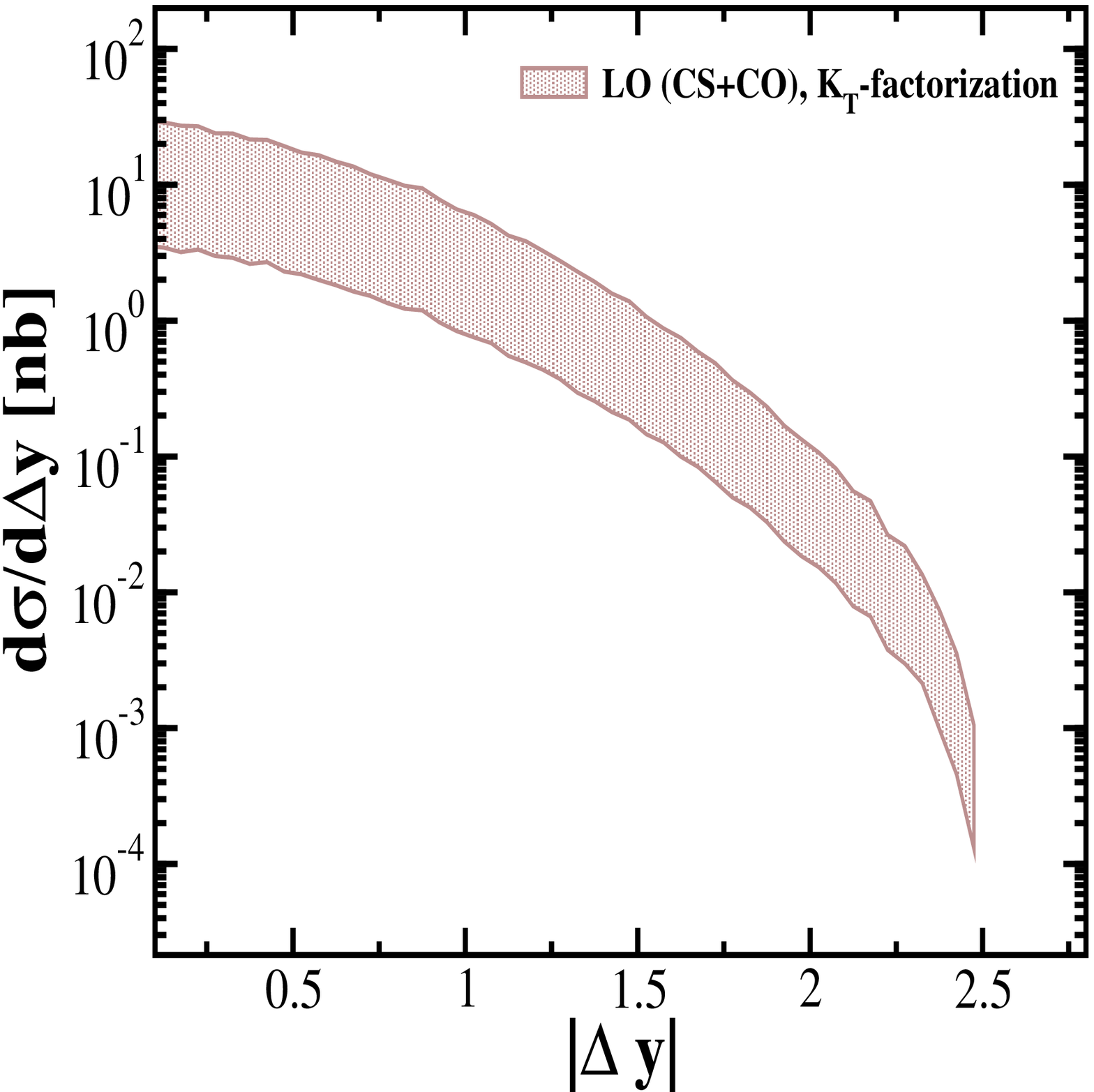}   
\includegraphics[width=8 cm] {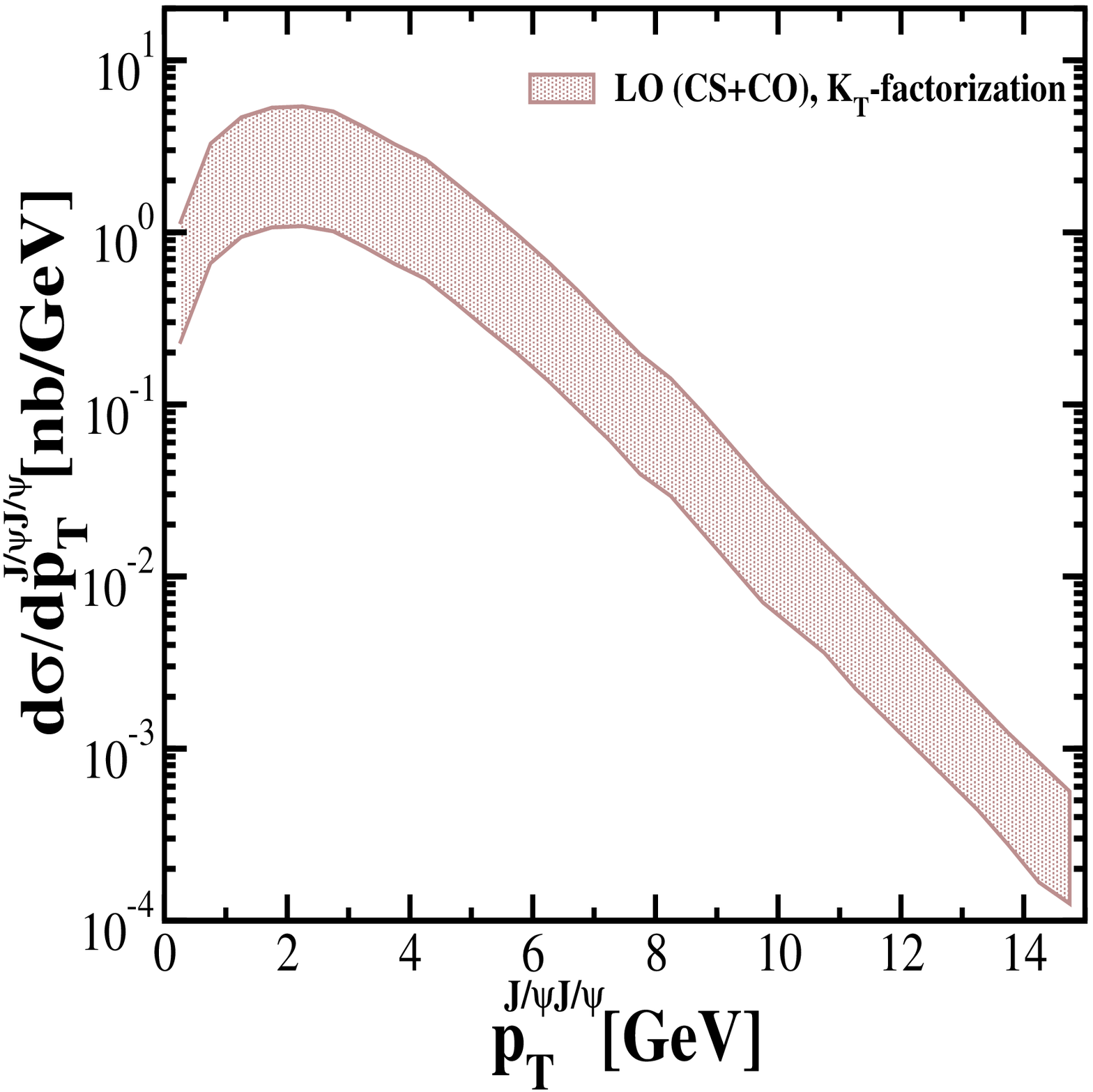}      
\includegraphics[width=8 cm] {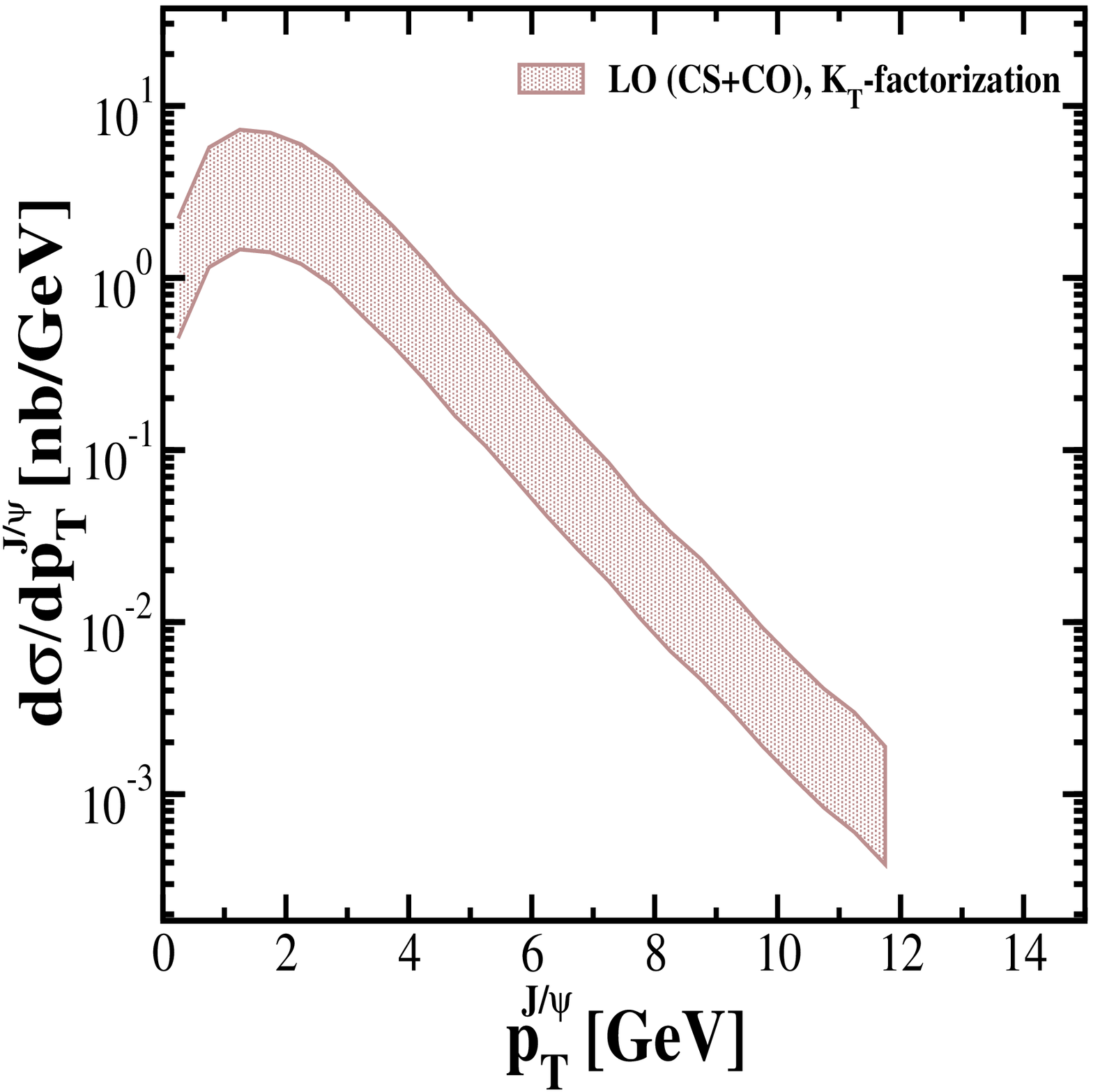}                            
\caption{Upper left: $M$-distribution of prompt double $J/\Psi$ hadroproduction measured by the LHCb collaboration \cite{lhcb} is compared to LO NRQCD predictions calculated in two different approaches: the $k_T$-factorization labeled by "LO (CS+CO), $k_T$-factorization" and the collinear factorization labeled by "LO (CS+CO), collinear fac."  by including CS and CO contributions. In upper left panel, we show the results obtained in the LO NRQCD $k_T$-factorization approach in the CS channel by taking two different unintegrated gluon density distributions (UGD) sets labeled by $A^{+}$ and  $A^{-}$.  We also show predictions for $|\Delta y|$-distribution (top-right panel), $p^{J/\Psi J/\Psi}_T$-distribution (lower-left panel)  and $p^{J/\Psi}_T$-spectra (lower-right panel) of prompt double $J/\Psi$ hadroproduction calculated in the NRQCD framework by including CS and CO contributions in the $k_T$-factorization approach. The band incorporates various theoretical uncertainties, see the text. The orange band (in upper left panel) labeled  by "LO (CS+CO), collinear fac." is taken from \cite{prl}.}
\label{f-lhcb}
\end{figure}

Here for the first time we confront the $k_T$-factorization based NRQCD results with the LHC data. The main advantage of the $k_T$-factorization is that one can incorporate the small-x dynamics in a minimal way via the unintegrated gluon density  $\mathcal{F}_g$ and its evolution. 
The unintegrated gluon densities absorb the effects of soft gluon resummation; 
this regularizes infrared divergences and makes our theory applicable to even small transverse momentum region.
At the LHC kinematics that we consider here, the Bjorken $x$ is quite small and therefore small-x physics is important. An approximate relation between the mass $M$ of the produced $J/\psi$-pair and 
the center-of-mass-energy squared $s$ of the colliding protons is $x_1 x_2  = M^2/s$. Therefore for
both the CMS and the LHCb kinematic coverage shown in Figs.\,\ref{f-lhcb},\ref{f-cms}, we have approximately  $x_{1,2} \leq 10^{-2}$.  
Note that in the $k_T$-factorization approach, the evaluation of Feynman diagrams is straightforward and follows 
standard QCD rules, with one reservation: in accordance with the  $k_T$
prescription \cite{sg}, the initial gluon spin density matrix is
taken in the form
$\overline{\epsilon_g^{\mu}\epsilon_g^{*\nu}}=k_T^\mu k_T^\nu/|k_T|^2,$
where $k_T$ is the component of the gluon momentum perpendicular to the
beam axis. In the collinear limit, when $k_T\to 0$, this expression 
converges to the ordinary 
$\overline{\epsilon_g^{\mu}\epsilon_g^{*\nu}}=-g^{\mu\nu}/2$,
while in the case of off-shell gluons it contains an admixture of
longitudinal polarization. 
We have checked that in the collinear limit we reproduce the results
of Refs. \cite{prl,Hmery,Scott,old1}. We also compare our results with those results coming from the collinear factorization which is based on a similar factorization given in  \eq{e1} by replacing the $\mathcal{F}_g$  convolution to a convolution with patron distribution functions (PDFs).

In the NRQCD factorization, in principle, $J/\Psi$ can also be produced by a branching processes of intermediate hadrons such as $\chi_{cJ}$ 
and $\psi^\prime$. The quantum numbers of $\psi'$ are identical to those of $J/\psi$, and therefore
 the production mechanism and all observables are also identical. Given that the
long-distance matrix elements for $J/\psi$ and $\psi'$ are proportional to the
respective leptonic decay widths, we have,   
$\langle O^{\psi'}_m\rangle \simeq (1/2) \langle O^{J/\psi}_m\rangle$. Therefore, we find that 
 $\psi'$ is produced with twice as small probability. Moreover,
 there is a $50\%$ branching fraction for $\psi'\to J/\psi+X$. Therefore, the feed-down 
 from $\psi'$ reduces to a multiplicative factor for the cross-section which can be incorporated as a pre-factor. 
Production of $\chi_c$ in combination with
 $J/\psi$ is forbidden by charge parity conservation (at LO). 
 Production of $\chi_c$ pairs is suppressed by their wave functions by two orders 
 of magnitude compared to $J/\psi$ pairs ($P$-waves versus $S$-waves). Therefore, this
 feed-down can safely be neglected. The production of $\chi_c$ pairs in the collinear factorization was considered in \cite{2chic}. 

\section{Numerical results and discussion}
\begin{figure}[t]                                       
\includegraphics[width=8 cm] {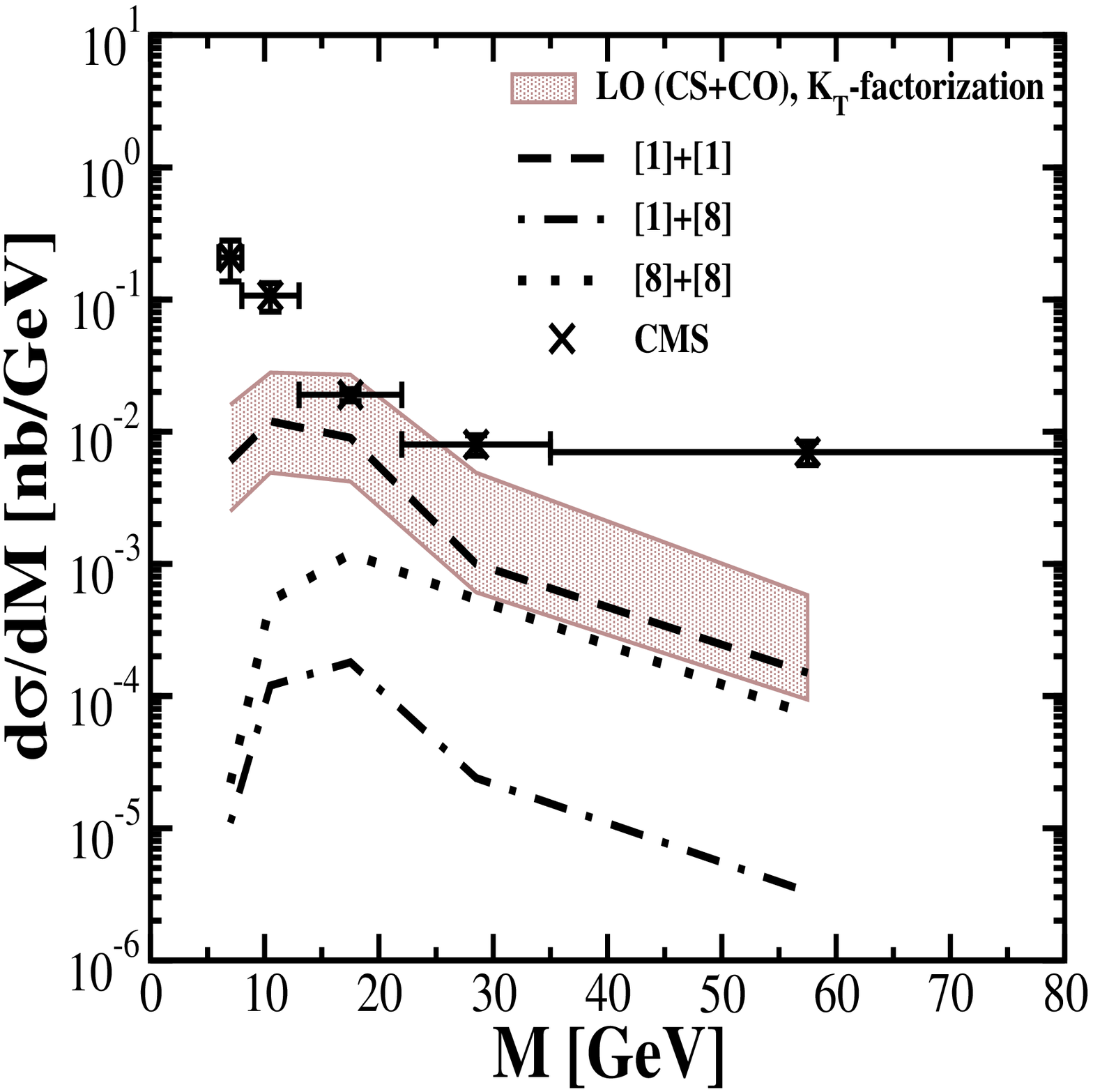}   
\includegraphics[width=8 cm] {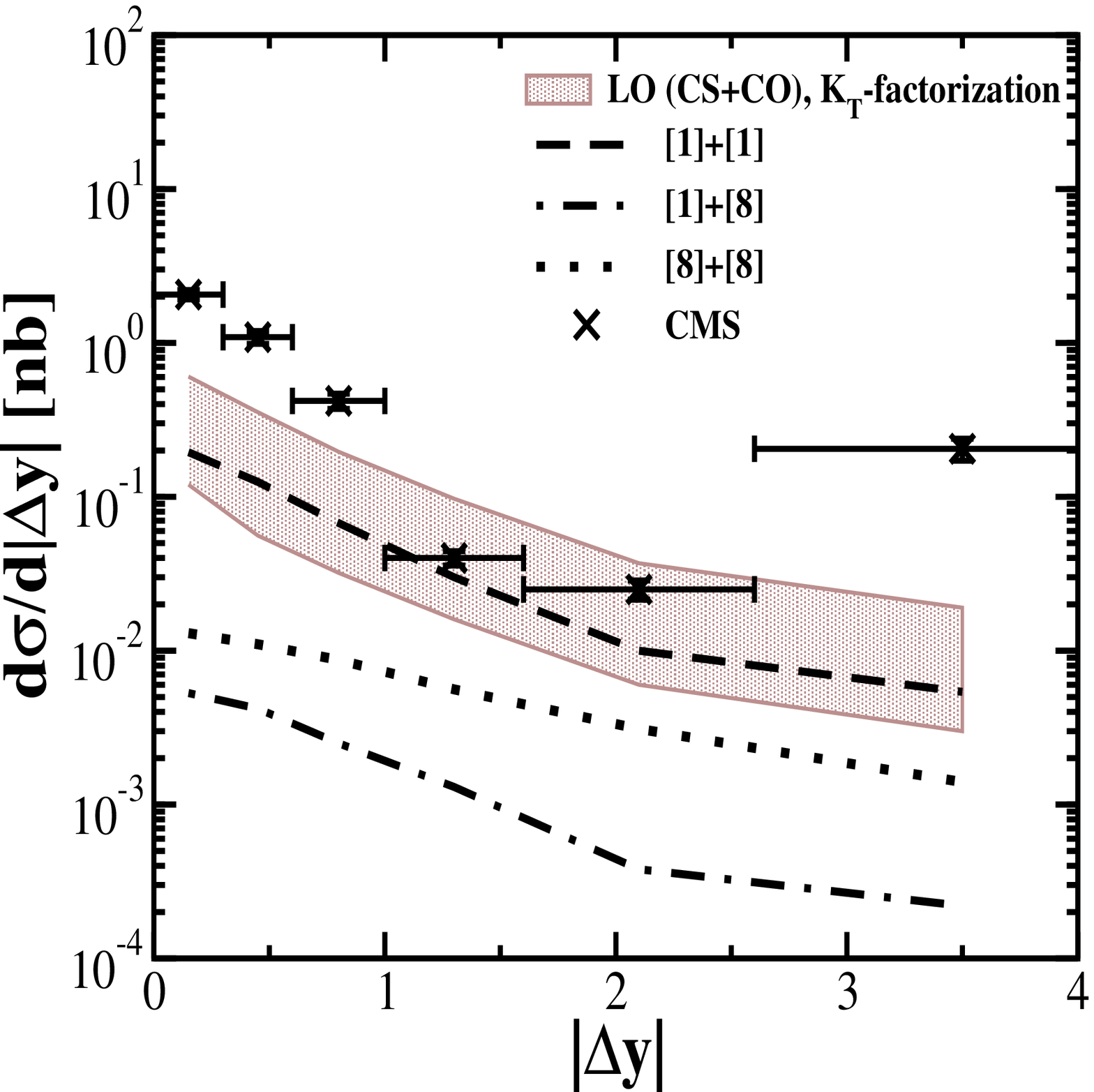} 
\includegraphics[width=8 cm] {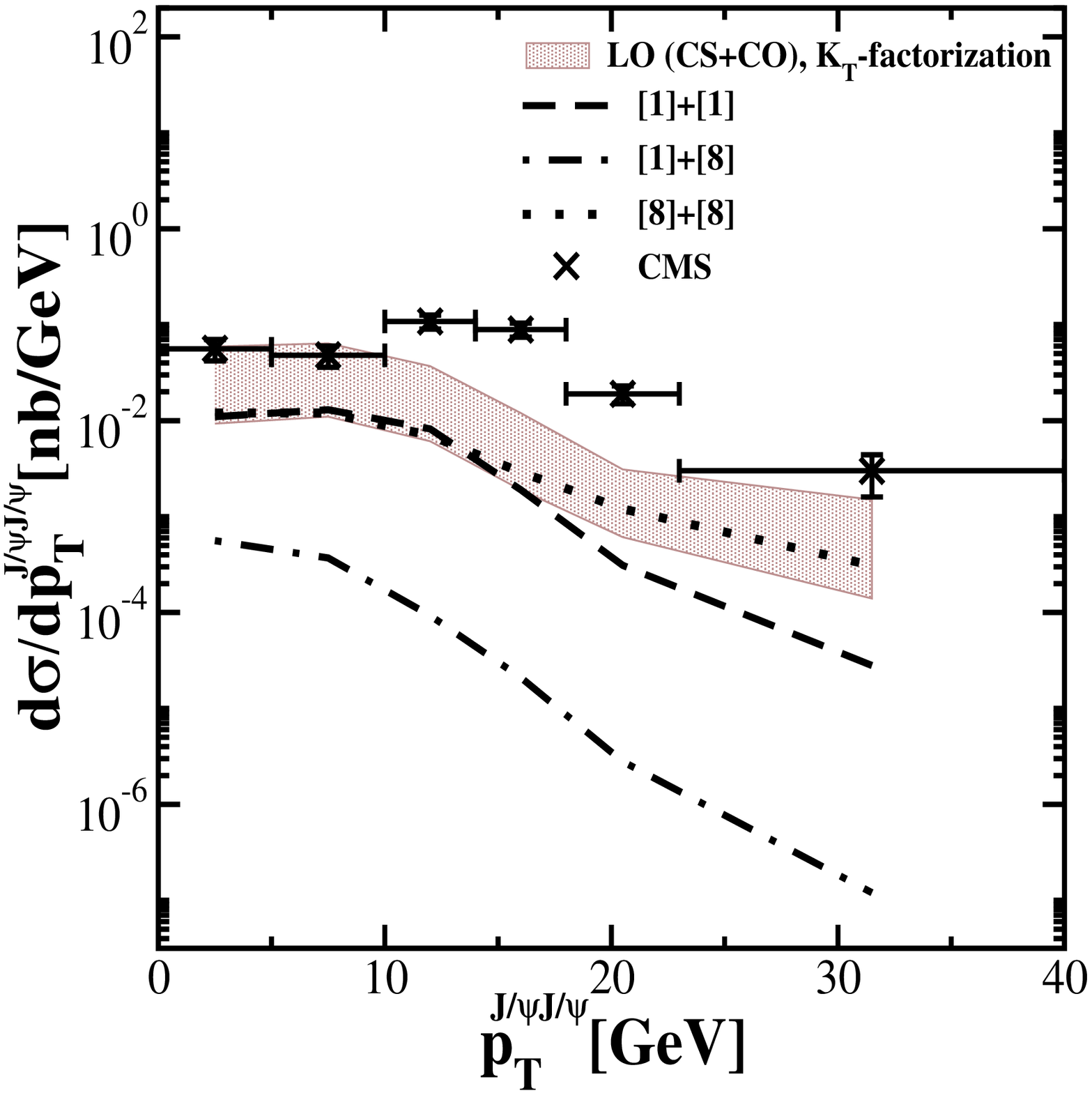}         
\caption{$M$-distribution (top left panel),  $|\Delta y|$-distribution (top right panel) and $p^{J/\Psi J/\Psi}_T$-distribution (lower panel) of prompt double $J/\Psi$ hadroproduction measured by the CMS collaboration \cite{cms} are compared to LO NRQCD predictions calculated in the $k_T$-factorization including CS and CO contributions. We also compare different contributions in the CS ([1]+[1]) and the CO ([1]+[8]), [8]+[8]) channels to the total cross-section and the experimental data.   }
\label{f-cms-0}
\end{figure}

\begin{figure}[t]                                       
\includegraphics[width=8 cm] {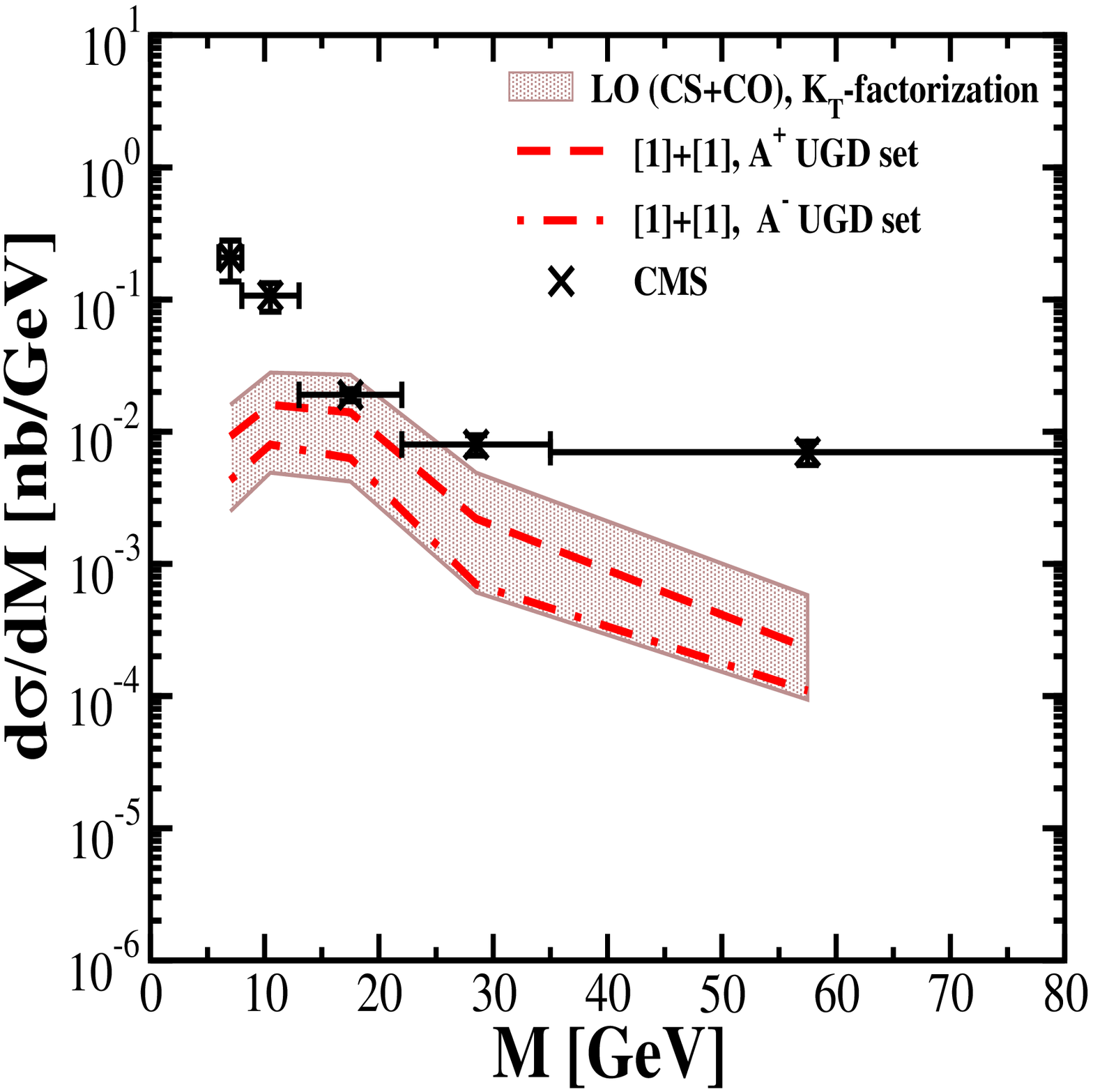}         
\includegraphics[width=8 cm] {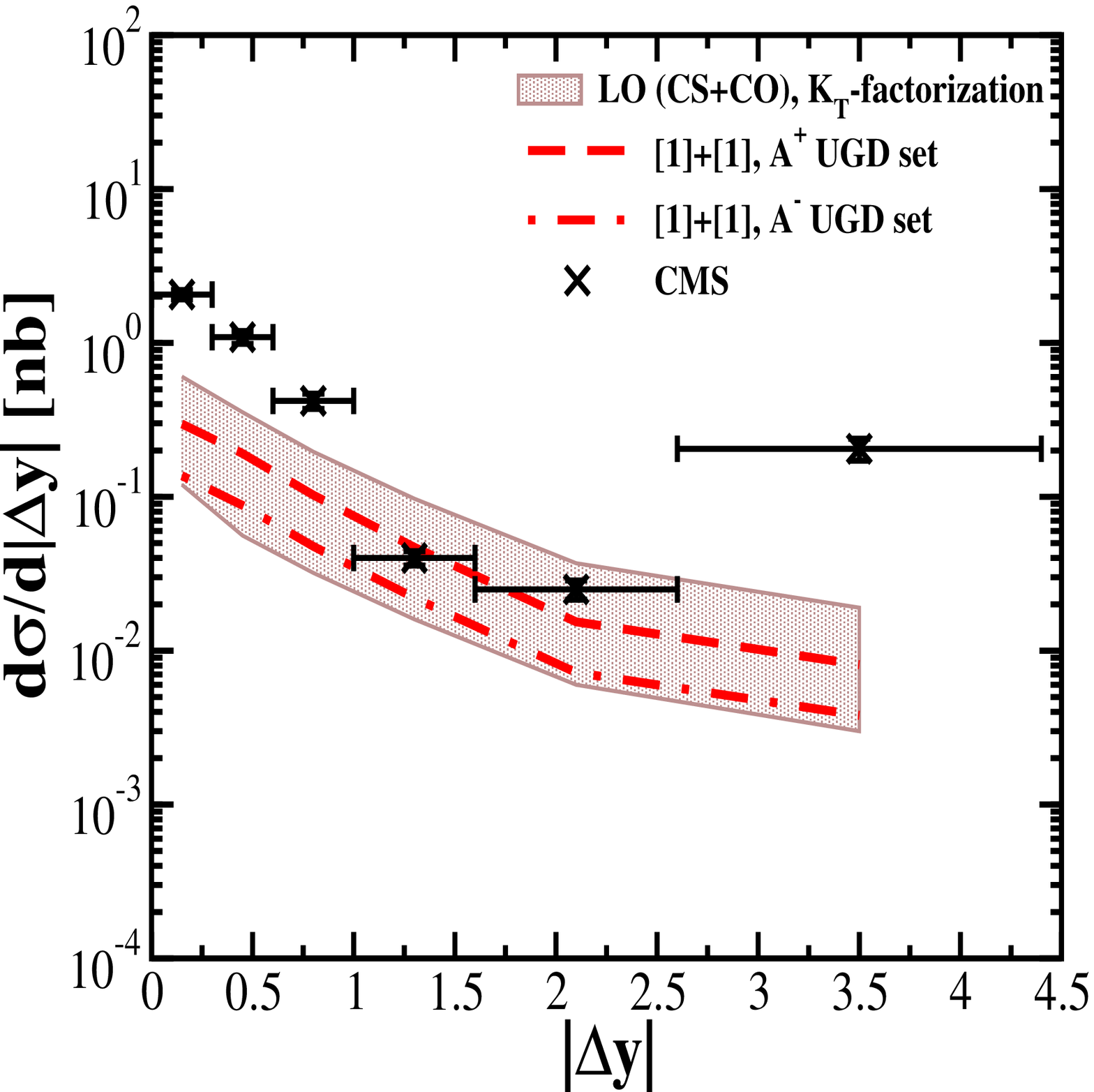}                          
\caption{$M$-distribution (left panel) and $|\Delta y|$-distribution (right panel) of prompt double $J/\Psi$ hadroproduction measured by the CMS collaboration \cite{cms} are compared to full LO NRQCD predictions calculated in the $k_T$-factorization. We also show the results obtained in the LO NRQCD $k_T$-factorization approach in the CS channel by taking two different unintegrated gluon density distributions (UGD) sets labeled by $A^{+}$ and  $A^{-}$. }
\label{f-cms-0-1}
\end{figure}

\begin{figure}[t]                                       
\includegraphics[width=8 cm] {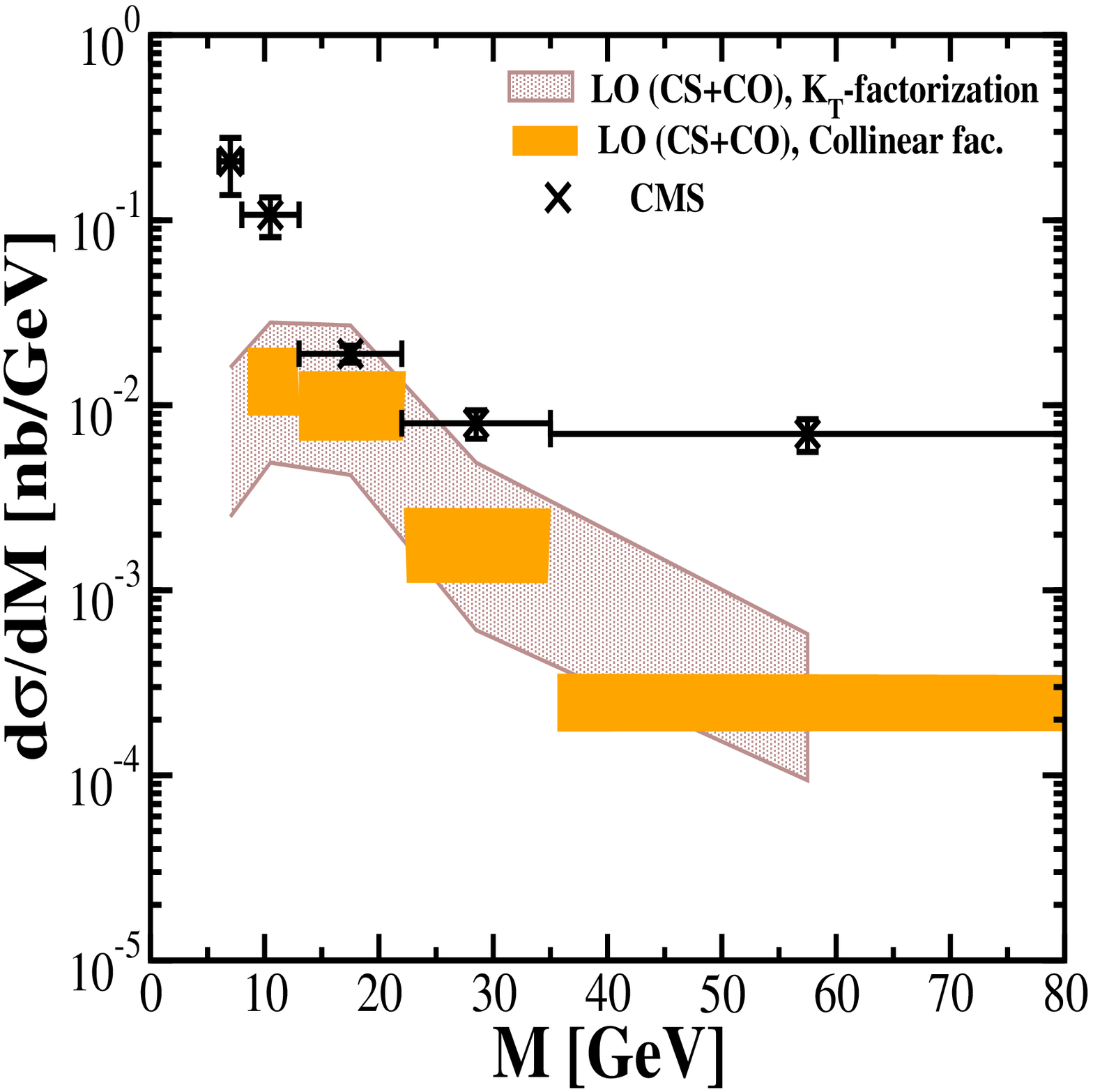}   
\includegraphics[width=8 cm] {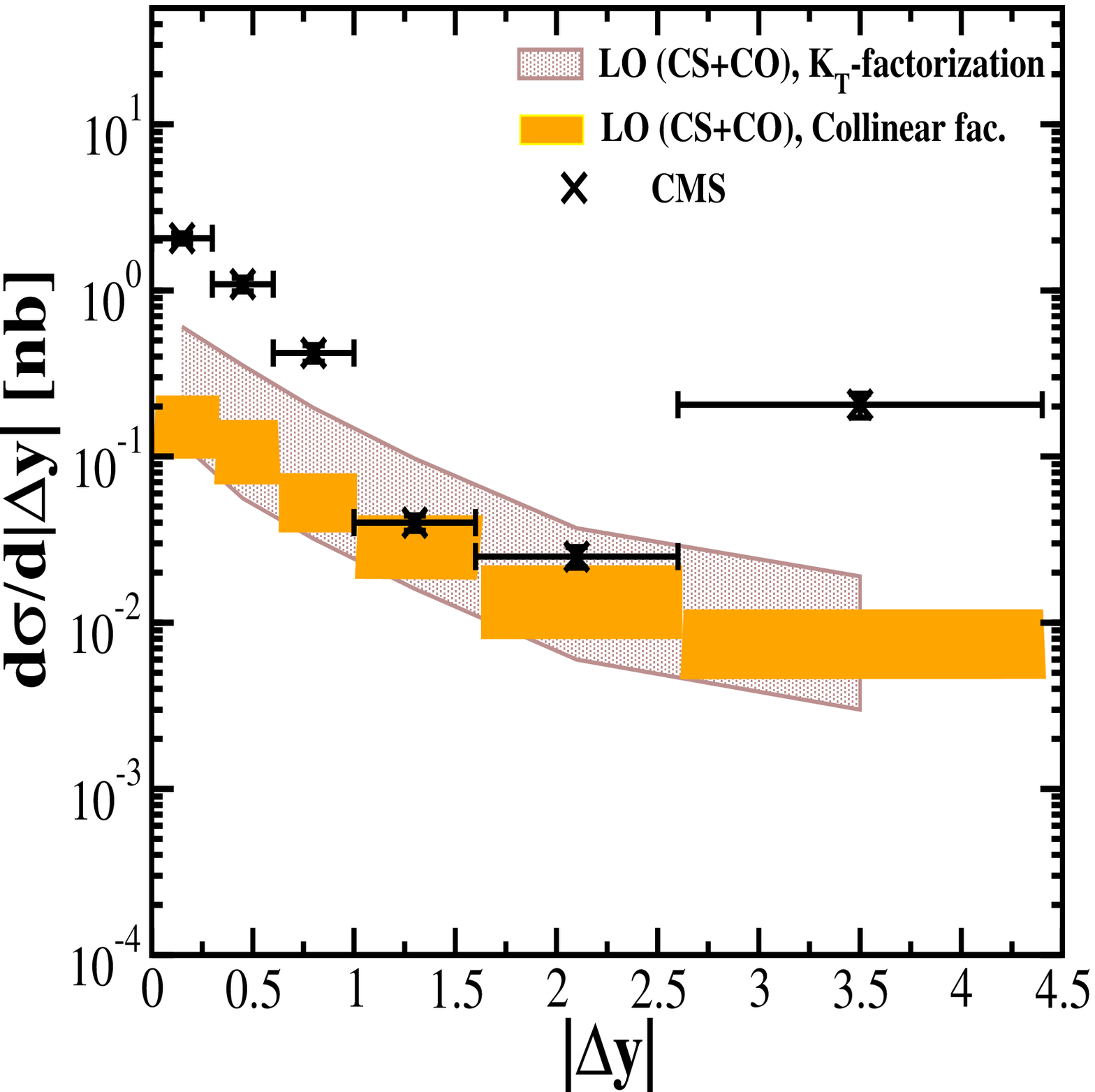}       
\caption{$M$-distribution (left panel) and $|\Delta y|$-distribution (right panel) of prompt double $J/\Psi$ hadroproduction measured by the CMS collaboration \cite{cms} are compared to full LO NRQCD predictions (including CS and CO contributions) calculated in two different approaches of the collinear factorization and the $k_T$-factorization. The notation of bands are the same as \fig{f-lhcb}. }
\label{f-cms}
\end{figure}

The parameter settings used in our numerical study is as follows.  Similar to the previous studies, we perform our computation at the LO in the fixed-flavor number scheme with three massless quark flavors and a charm quark with a mass approximately taken to be half of $J/\Psi$ mass $m_c \approx m_{J/\Psi}/2$. Note that there is large uncertainties on the extracted value of charm mass from the current data for the charm structure functions and diffractive processes at HERA at the small-x region \cite{charm-hera}. Following the standard pQCD practice at LO, we assume the factorization and the renormalization scale to be equal. We use the LO running coupling $\alpha_s(\mu_R^2)$  with a renormalization scale $\mu_R^2=\hat{s}/4$ where $\hat{s}$  is the partonic subprocess
invariant energy  squared. Note that this choice of $\mu_R$ is Lorentz invariant and symmetric with respect to $J/\Psi$ pair. 
In the theoretical bands shown in Figs.\,\ref{f-lhcb}-\ref{f-cms} we incorporate the uncertainties associated to our freedom to choose different factorization scale by varying  $\mu_R$ in a range of $\hat{s}/8<\mu_R^2<\hat{s}/2$.   We employed the unintegrated gluon density (UGD) $\mathcal{F}_g$ obtained from the CCFM evolution \cite{ccf} constrained by data on the structure function. The CCFM evolution equation matches 
to the (LO) DGLAP evolution equation at moderate $x$ and it embodies BFKL dynamics at low x \cite{ccf}. The free parameters of the UGD was obtained from a fit to the HERA structure functions data with $x<5.10^{-3}$ and $Q^2>4.5\,\text{GeV}^2$ \cite{Jung}. Note that the current HERA data alone cannot uniquely determine the free parameters of the UGDs. We incorporate the uncertainties associated to our freedom to choose different UGDs by incorporating in the theoretical bands shown in Figs.\,\ref{f-lhcb}-\ref{f-cms} the effect of employing different UGD sets ($A^0$, $A^+$ and $A^-$ sets given in Ref.\,\cite{Jung}). 

The only free parameter of the long-distance matrix element $ \langle O^{J/\Psi}_1 (^3S_1^{[1]})\rangle$ in \eq{e1} in the CS channel is the $J/\Psi$ radial wavefunction  at the origin of coordinate  space $\mathcal{R}_{J/\Psi}(0)$. The parameter $\mathcal{R}_{J/\Psi}(0)$ is determined via  a fit to the leptonic decay width \cite{leptonic} with an extracted value of $|\mathcal{R}_{J/\Psi}^2(0)|=0.8\, \text{GeV}^3$. The nonperturbative long-distance color-octet matrix elements are taken from Ref. \cite{ChoLei}:
$\langle O^{J/\Psi}_1 (^3S_1^{[8]})\rangle=1.2\times 10^{-2}\,\text{GeV}^3$
$\langle O^{J/\Psi}_1 (^3P_0^{[8]})\rangle=1.6\times10^{-2}\,\text{GeV}^3$.

At the LHCb, the $J/\Psi$ pair production are sampled at $\sqrt{s}=7$ TeV in the following kinematic phase space for the individual $J/\Psi$ with transverse momentum $p_T^{J/\psi}$ and rapidity $y^{J/\psi}$ \cite{lhcb}: 
\begin{equation}\label{c1}
0<p_T^{J/\psi}[\text{GeV}]<10\,\,\,\,\,\text{for}\,\,\,\,\, 2<y^{J/\psi}<4.5. 
\end{equation}
Note that in this paper, the transverse momentum of a $J/\Psi$-pair and an individual $J/\Psi$ are denoted by $p^{J/\Psi J/\Psi}_T$ and $p_T^{J/\psi}$, respectively.
The CMS experiment samples the $J/\Psi$ pair production at the same energy $\sqrt{s}=7$ TeV but complementary to the LHCb kinematics, with coverage to higher transverse momentum of a pair at more central rapidity region. The CMS experiment provides access to $p^{J/\Psi J/\Psi}_T$ measurments above $15$ GeV.  The CMS measured the cross-section in a phase space defined by the individual $J/\Psi$ \cite{cms}: 
\begin{eqnarray}\label{c2}
&&p_T^{J/\psi}[\text{GeV}]>6.5\,\,\,\,\,\text{for}\,\,\,\,\, |y^{J/\psi}|<1.2 \nonumber\\ 
&&p_T^{J/\psi}[\text{GeV}]>6.5\to 4.5\,\,\,\,\,\text{for}\,\,\,\,\, 1.2<|y^{J/\psi}|<1.43 \nonumber\\ 
&&p_T^{J/\psi}[\text{GeV}]>4.5\,\,\,\,\,\text{for}\,\,\,\,\, 1.43<|y^{J/\psi}|<2.2. \
\end{eqnarray}
For a comparison with the LHCb and CMS data, in our numerical calculation we impose the same kinematic constrains as employed by the LHCb and the CMS experiments, given in Eqs.\,(\ref{c1},\ref{c2}).  

The theoretical uncertainties are shown by a band in Figs.\,\ref{f-lhcb}-\ref{f-cms}. In the $k_T$-factorization approach, main uncertainties come from our freedom to choose different factorization/renormalization scale and different UGDs. To highlight the effect of different UGDs, we also show in Figs.\,\ref{f-lhcb},\ref{f-cms-0-1}, the results obtained by two different sets of the UGD \cite{Jung} labeled by $A^+$ and $A^-$ UGD sets in the CS channel. It is seen that the uncertainties associated to the UGDs are rather large. Note that both the collinear factorization and the $k_T$-factorization bands shown in Figs.\,\ref{f-lhcb}-\ref{f-cms} include all feed-down channels at LO.

In \fig{f-lhcb}, we first confront the LHCb data for the differential cross-section for $J/\Psi$ pairs as a function of the invariant mass $M$ of  the $J/\Psi$-pair system with LO NRQCD results calculated in two different formalisms of the collinear factorization and the $k_T$-factorization. 
Note that soft gluon radiations (or initial-state radiation) incorporated in the $k_T$-factorization formalism are important and drastically modifies the shape of $p^{J/\Psi J/\Psi}_T$,  $p^{J/\Psi}_T$ and $M$  distribution at low transverse momentum. It is generally seen that within the error bands, the LO NRQCD results obtained in the $k_T$-factorization are consistent with the LHCb data over entire range of $M$ shown in \fig{f-lhcb}.  At high  $M$, both the collinear and the $k_T$-factorization results are consistent with each other. Note at the LHCb kinematics shown in  \fig{f-lhcb}, the CS contribution dominate and the CO contributions are negligible. In order to highlight this effect, in  \fig{f-lhcb} (top-left panel), we also compare with the results obtained by only taking into account the CS channel.

In order to further test the NRQCD framework at the LHC, in \fig{f-lhcb}, we show our predictions for the rapity distribution of double $J/\Psi$ production as a function of rapidity difference between two $J/\Psi$  (top right panel), $p^{J/\Psi J/\Psi}_T$-distribution (lower left)  and $p^{J/\Psi}_T$-spectra (lower right) of prompt double $J/\Psi$ hadroproduction at LO in the $k_T$-factorization approach employing the same kinematic phase space as taken at the LHCb (given in \eq{c1}). Again similar to upper-left panel in \fig{f-lhcb}, the band labeled by "LO(CS+CO), $k_T$-factorization" includes both CS and CO contributions and also various theoretical uncertainties outlined above.

One can expect that at very large transverse momentum $p_T$, the dominant contribution comes
from the gluon-gluon scatterings subprocess $g+g\to g^*+g^*$ followed by
gluon fragmentation $g^*\to J/\psi$. This mechanism is generally suppressed
by the relatively low values of color-octet matrix elements in comparison
with the color-singlet one. However, the specific $p_T$ dependence of the gluon-gluon scatterings process is quite different in the CS and the CO channels, namely we have $d\sigma/dp_t\propto 1/p_T^4$ in the CO fragmentation channel, while for the color-singlet production one
has $d\sigma/dp_t\propto 1/p_T^8$ (see table I in \cite{prl}). Therefore, one expects that the CO channels only become important at high transverse momentum. This can be seen in Figs.\,\ref{f-lhcb},\ref{f-cms-0}. 

In \fig{f-cms-0}, we compare various contributions of the CS and the CO channels to the $M$-distribution (top-left panel), the $|\Delta y|$-distribution (top-right panel) and $p^{J/\Psi J/\Psi}_T$-distribution (lower panel) of prompt double $J/\Psi$ hadroproduction. All theoretical curves in \fig{f-cms-0} are obtained in the $k_T$-factorized LO NRQCD. In \fig{f-cms-0}, we also show the CMS data \cite{cms}. We recall that the CMS data shown in \fig{f-cms-0} are taken in the kinematic phase space given in \eq{c2}. In \fig{f-cms-0}, for simplicity of notation,  we denote the CS and the CO channels for $c\bar{c}$ production with $^3S_1^{[1]}=[1]$  and $^1S_0^{[8]}+^3S_1^{[8]}+^3P_0^{[8]}+^3P_1^{[8]}+^3P_2^{[8]}=[8]$, respectively. It is generally seen in \fig{f-cms-0} that the CO channels become only important at high $M$ and high transverse momentum. It is also seen that at high transverse momentum we have $[8]+[8]>[1]+[8]$. Given rather large theoretical uncertainties at LO, it is not possible to make any firm conclusion if the CO contribution is essential at the CMS kinematics at intermediate transverse momentum.  Nevertheless, it is seen in \fig{f-cms-0} that at high $p^{J/\Psi J/\Psi}_T$, the contribution of the CO ([8]+[8]) channel becomes larger than the CS([1]+[1]) one.  

In order to highlight rather large theoretical uncertainties due to the unintegrated gluon density distributions (UGDs), in \fig{f-cms-0-1}, we compare the CMS data with the results obtained in the  $k_T$-factorized LO NRQCD in the CS channel by taking two different UGD sets labeled by $A^{+}$ and  $A^{-}$.

In \fig{f-cms}, we compare the $M$-distribution (left panel) and the $|\Delta y|$-distribution (right panel) of prompt double $J/\Psi$ hadroproduction measured by the CMS collaboration \cite{cms} with full LO NRQCD predictions calculated in two different approaches of the collinear factorization and the $k_T$-factorization (including both the CS and the CO contributions). The notation and description of theoretical bands are similar to \fig{f-lhcb}. It is seen that within the theoretical uncertainties, the results obtained in the $k_T$-factorization and the collinear factorization approaches are consistent in the CMS kinematics shown in \fig{f-cms}. It is generally seen that full LO NRQCD results based on both the $k_T$-factorization and the collinear factorization formalisms, significantly underestimate the CMS data in most of $M$, $p^{J/\Psi J/\Psi}_T$ and $|\Delta y|$ bins. Note that $M$,  $p^{J/\Psi J/\Psi}_T$ and $|\Delta y|$  are related, and a large  $p^{J/\Psi J/\Psi}_T$ corresponds to a large $M$. It is generally seen in \fig{f-cms} that the discrepancy between theoretical results at LO and the CMS data (within error bars) can be as large as about two orders of magnitude at large $M$ (and/or large $p^{J/\Psi J/\Psi}_T$). Taking the CMS data at face value,  the flatness of the data at large value of $M$ and $|\Delta y|$, and raise of the $|\Delta y|$ distribution at large  $|\Delta y|$ indicate a large missing anamolous contribution beyond the full LO approximation, see also Ref. \cite{cms3}. 

\begin{figure}[t]                                       
\includegraphics[width=8.5 cm] {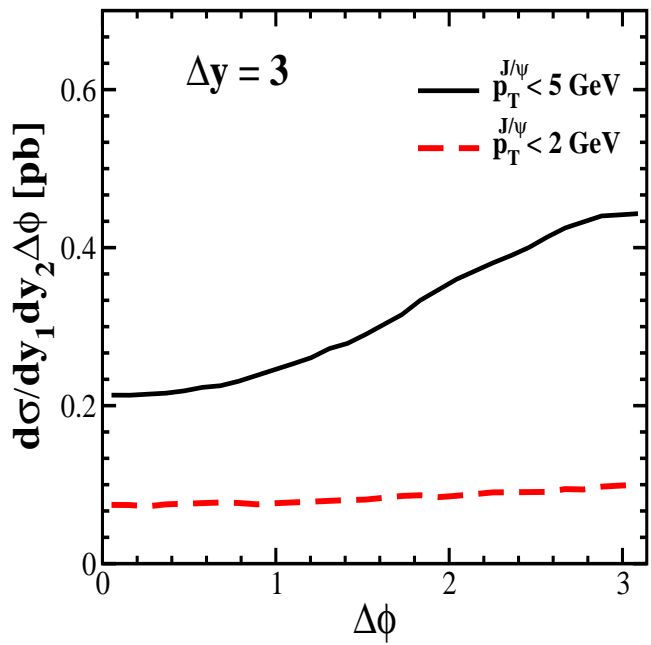}   
\caption{ The prompt double $J/\Psi$ cross-section as a function of angle between two $J/\Psi$ at the LHC energy $8$ TeV. The rapidities of two $J/\Psi$ are taken at $y_1^{J/\Psi}= 1$ and  $y_2^{J/\Psi}= 4$ (with $\Delta y = 3$). The solid and dashed curves are obtained by integrating over the individual transverse momentum of $J/\Psi$ for  $p_T^{J/\Psi}< 5$ GeV and $p_T^{J/\Psi}< 2$ GeV, respectively. }
\label{f-phi1}
\end{figure}

In \fig{f-phi1}, we show the prompt double $J/\Psi$ cross-section as a function of angle $\Delta \phi$ between two $J/\Psi$ at the LHC energy $8$ TeV for a case that the rapidity interval between two  $J/\Psi$ is taken $\Delta y = 3$ (with $y_1^{J/\Psi}=1$ and $y_2^{J/\Psi}=4$) and transverse momenta of each of $J/\Psi$ was integrated over $p^{J/\Psi}_T<5$  (solid line) and $p^{J/\Psi}_T<2$  (dashed line). The result shown in  \fig{f-phi1} was obtained by employing $A^{0}$ UGD set and a fixed factorization/renormalization scale $\mu_F^2 = \mu_R^2 = \hat{s}/4$. It is seen in \fig{f-phi1}  that in both cases of different transverse momentum bins, there is no enhancement at near-side $\Delta \phi\approx 0$ while the way-side $\Delta \phi\approx \pi $ correlations are enhanced at larger $p^{J/\Psi}_T$ bins due to standard back-to-back kinematics.

Although the standard double parton interactions contributions  \cite{sub} (not considered here) are important at large rapidity interval, and may change the normalization of the total cross-section, but such contributions will not modify the flatness of 
the correlations in angle between two $J/\Psi$ at near-side. We have also checked that the inclusion of subleading $\alpha_s^6$  corrections (due to gluon exchanges to the box-diagrams)  which dominate at very large rapidity interval \cite{sub}, will not change the flatness of the near-side correlation for $J/\Psi$-pairs.  Note that such subleading $\alpha_s^6$  corrections are negligible at $\Delta y = 3$ employed in \fig{f-phi1}. We recall that at the same rapidity and transverse momentum coverage considered here, di-hadron correlations exhibit a ridge-like structure in proton-proton collisions at the LHC in high-multiplicity event selections.  Note that the experimental data for $J/\Psi$-pairs correlations in proton-proton and proton-nucleus collisions are not yet available.  Our results show that at LO in NRQCD framework, there is no a ridge-like behaviour (a second local maximum at the near-side) for prompt double $J/\Psi$  production in proton-proton collisions at the LHC.

\section{Conclusion}
In this paper we confronted the full LO NRQCD results obtained in two approaches of the collinear factorization and the $k_T$-factorization with the recent LHC data for prompt double  $J/\Psi$ production in proton-proton collisions. We showed that the LO $k_T$-factorized NRQCD results are consistent with the LHCb data. Here we provided various predictions at the LHCb kinematic phase space which can further test the NRQCD formalism, see \fig{f-lhcb}.  We quantified various theoretical uncertainties in the NRQCD formalism and showed that the CMS data cannot be described at LO with about a factor of 10 discrepancy, see \fig{f-cms}. It remains to be seen if full next-to-leading-order (NLO) corrections can fill such a large gap between LO results and the CMS data, see also Refs.\,\cite{cms0,cms1,cms2,cms3}. Although the relative importance of the NLO corrections is process-dependent, we recall that the NLO corrections enhance the LO NRQCD cross-section for single inclusive $J/\Psi$ production by a factor about 10 \cite{ref-nlo}, see also Refs.\,\cite{nlo-s,nlo-sj}.    
Note that in this paper, we only focused on the single parton scattering (SPS) contributions. The contribution of double parton scattering (DPS) \cite{sub,dps}  and generally multiparton interactions \cite{mpi} can be important for double $J/\Psi$ production at small-x region at forward rapidities. The D0 collaboration \cite{d0}, for the first time, separated the double $J/\Psi$ production cross section at $\sqrt{s}=1.96$ TeV into contributions due to single and double parton scatterings, and found that $(46\div 22)\%$ of cross-section is due to DPS. Although, the relative importance of the DPS cannot be unambiguously quantified till full NLO correlations are known, it is hard to believe that one order of magnitude discrepancy seen between theory and CMS data at about midrapidity can be accounted for by the DPS. 

Within the $k_T$-factorized NRQCD approach for prompt double $J/\Psi$ production, we quantified various theoretical uncertainties including  the uncertainties associated to the unintegrated gluon density which was previously neglected. Within these theoretical uncertainties and rather large experimental error bars, with only the current LHC data for prompt double  $J/\Psi$ production, one cannot conclusively pin down the importance of the CO contribution. In particular, we showed that the LHCb data can be described by only taking into account the CS channel. We showed that 
the CO contributions become important at high $M$ and high $p^{J/\Psi J/\Psi}_T$ (at the CMS kinematics, see \fig{f-cms-0}). Nevertheless, in order to further substantiate the importance of the CO contribution, it is indispensable to complete the full NLO analysis of prompt double $J/\Psi$ production by including all CO channels.  

We also investigated the long-range in rapidity azimuthal angle correlations between $J/\Psi$ pairs in the NRQCD formalism. We found that there is no near-side ridge-like structure up to subleading $\alpha_s^6$  accuracy in this formalism. It is of great interest to see whether the inclusion of higher order corrections, the DPS effect and in particular density effect at high multiplicity events, and in general  small-x dynamics due to gluon saturation or color-glass-condensate physics, can change this outcome. Note that there are growing evidence that the gluon saturation physics (which was not considered here) is important in small-x region in proton-proton collisions  at the LHC, see e.g. Refs.\,\cite{ridge0,ridge1,ridge2,ridge3,ridge33,pp-amir}. These are important issues which are beyond the scope of this paper and certainly deserve separate studies.

\begin{acknowledgments}
The work of A.H.R. is supported in part by Fondecyt grant 1110781, 1150135 and Conicyt C14E01.
\end{acknowledgments}


\end{document}